\documentclass[11pt]{article}
\usepackage{epsfig}
\usepackage{epstopdf}
\usepackage{amsmath}
\usepackage{amssymb}
\usepackage{mathrsfs}
\usepackage{graphicx}
\newsavebox{\DSLASH}
\sbox{\DSLASH}{$D$\hspace{-2.5mm}/}
\newcommand{\DS}{\usebox{\DSLASH}}

\begin{document}

\vspace{4cm}
\begin{center}
{\Large\bf{Single top quark production in $t$-channel at the LHC in Noncommutative Space-Time}}\\

\vspace{1cm}
{\bf Seyed Yaser Ayazi$^{\dagger}$, }
{\bf Sina Esmaeili$^{\ddagger}$,}
{\bf  Mojtaba Mohammadi Najafabadi$^{\dagger}$ }  \\
\vspace{0.5cm}
{\sl ${\ ^{\dagger}}$  School of Particles and Accelerators, \\
Institute for Research in Fundamental Sciences (IPM) \\
P.O. Box 19395-5531, Tehran, Iran}\\
and \\
{\sl ${\ ^{\ddagger}}$  Science and Research Branch, Islamic Azad University}\\

\vspace{2cm}
 \textbf{Abstract}\\
 \end{center}

We study the production cross section of the $t$-channel single top quark at the LHC in
the noncommutative space-time. It is shown that the deviation of the $t$-channel single top
cross section from the Standard Model value because of noncommutativity is significant when
$|\vec{\theta}| \gtrsim 10^{-4}$ GeV$^{-2}$.
Using the present experimental precision in measurement of the $t$-channel cross section, we apply
upper limit on the noncommutative parameter. When a single top quark decays, there is a significant amount of angular correlation, in the top quark rest frame
between the top spin direction and the direction of the charged lepton momentum from its decay. We study the effect of noncommutativity on the
spin correlation and we find that depending on the
noncommutative scale, the angular correlation can enhance considerably. Then, we provide limits on the noncommutative scale
for various possible relative uncertainties on the spin correlation measurement. 
\\

PACS number(s): 14.65.Ha

\newpage

\section{Introduction}

The study of single top quark processes at hadron colliders provides
the opportunities to investigate the electroweak properties of the top
quark, direct measurement of the $V_{tb}$ CKM matrix element, and more importantly it provides
the possibility to search for new physics.
The standard model (SM) has been found to be in a good agreement with the present experimental measurements
in many of its aspects. In the framework of the SM,
top quark is the heaviest particle with the mass at the
order of the electroweak symmetry breaking scale,
$v \sim$ 246 GeV.
This large mass might be a hint that the top quark plays an essential
role in the electroweak symmetry breaking. On the other
hand, the reported experimental data from Tevatron and LHC on the
top quark properties are still limited and no significant
deviations from the standard model predictions has been
observed yet.

Top quarks are mainly produced through two independent
mechanisms at hadron colliders: The main production mechanism is via strong interactions
where top quarks are produced in pair ($gg\rightarrow t\bar{t},q\bar{q}\rightarrow t\bar{t}$) \cite{werner}.
The production cross section of $t\bar{t}$ at 7 TeV center-of-mass energy at the LHC is 157 pb at next to leading order \cite{ttbar}.
Top quark can be produced singly via electroweak interaction. It occurs through three different processes
: $t$-channel (the involved $W$-boson is space-like, $ub\rightarrow dt$), $s$-channel
(the involved  $W$-boson is time-like, $u\bar{d}\rightarrow \bar{b}t$)
and $tW$-channel (the involved $W$-boson is real, $gb\rightarrow W^{-}t$). The $t$-channel with the cross section of 60 pb
is the largest source of single top at the LHC \cite{werner}.

The cross sections of $t\bar{t}$ and single top production,
the top quark mass,
the helicity of $W$ boson in top decay, the search for
flavor changing neutral current, and many other properties of the top
quark have been already studied \cite{werner},\cite{top}.
However, it is expected that top quark properties such as
single top quark cross section measurement are going to be
measured with high precision at the LHC
due to very large statistics \cite{werner}.

The space-time noncommutativity is a generalization
of the usual quantum mechanics and quantum field theory which may
describe the physics at short distances of the order of the
Planck length, since the nature of the space-time could change at
these distances. There are motivations coming from string theory, quantum
gravity, Lorentz breaking \cite{Douglas1},\cite{Douglas2},\cite{Ardalan1},\cite{Ardalan2})
to construct models on noncommutative space-time.
The noncommutativity in space-time can be
described by a set of constant c-number parameters
$\theta^{\mu\nu}$ or equivalently by an energy scale
$\Lambda_{NC}$ and dimensionless parameters $C^{\mu\nu}$:

\begin{eqnarray}
[\hat{x}_{\mu},\hat{x}_{\nu}] = i\theta_{\mu \nu} = \frac{i}{\Lambda^{2}_{NC}}C_{\mu\nu}
= \frac{i}{\Lambda^{2}_{NC}}\left(
\begin{array}{cccc}
0 & -E_{1} & -E_{2} & -E_{3} \\
E_{1} & 0 & -B_{3} & B_{2}   \\
E_{2} & B_{3} & 0 & -B_{1}   \\
E_{3} & -B_{2} & B_{1} & 0  \\
\end{array}
\right)
\end{eqnarray}
where $\theta_{\mu \nu}$ is a real anti-symmetric tensor which has the
dimension of $[M]^{-2}$. Dimensionless
electric and magnetic parameters $(\vec{E},\vec{B})$ have been defined for
convenience. It is notable that a space-time noncommutativity,
$\theta_{0i}\neq 0$, might cause some problems with unitarity
and causality \cite{Gomis},\cite{Chaichian}. It has been shown
that the unitarity can be satisfied for the case of
$\theta_{0i}\neq 0$ provided that $\theta^{\mu \nu}\theta_{\mu
\nu} > 0$ \cite{kostelecky}. However for simplicity, in this
article we take $\theta_{0i} = 0$ or equivalently $\vec{E} = 0$.

One can obtain a noncommutative version of an ordinary field theory
by replacing all ordinary products among fields with Moyal $\star$
product defined as \cite{review}:
\begin{eqnarray}
(f\star g)(x) &=& \exp\left(\frac{i}{2}\theta^{\mu
\nu}\partial_{\mu}^{y}\partial_{\nu}^{z}\right)f(y)g(z)\bigg\vert_{y=z=x}\\
\nonumber &=&
f(x)g(x)+\frac{i}{2}\theta^{\mu\nu}(\partial_{\mu}f(x))(\partial_{\nu}g(x))+O(\theta^{2}).
\end{eqnarray}

The approach to the noncommutative field theory based on the Moyal
product and Seiberg-Witten maps allows the generalization of the
standard model to the case of noncommutative space-time, keeping
the original gauge group and particle content \cite{SW},\cite{Madore1},\cite{Madore2},\cite{Madore3},\cite{Madore4},\cite{Reuter}.
Seiberg-Witten maps relate the noncommutative gauge fields and
ordinary fields in commutative theory via a power series
expansion in $\theta$. Indeed the noncommutative version of the
Standard Model is a Lorentz violating theory, but the Seiberg
Witten map shows that the zeroth order of the theory is the
Lorentz invariant Standard Model. The effects of noncommutative
space-time on some rare decay, collider processes, leptonic decay
of the $W$ and $Z$ bosons and additional phenomenological results
have been presented in \cite{OHL},\cite{Haghighat},\cite{Schupp},\cite{Martin1},\cite{mojtaba},\cite{Josip1},\cite{Josip2},\cite{Melic},\cite{Iltan},\cite{arfaei},\cite{josip3},\cite{hewett}
and some limits have been set on noncommutative scale.

In this article, we calculate the contributions that the
$t$-channel single top quark cross section receive from the
noncommutativity in space-time at the LHC in the center-of-mass
energy of 7 TeV. Then, we estimate a bound on the
noncommutative parameter $\theta$ by comparing the
recent measurement of the CMS collaboration of the
$t$-channel cross section with the theoretical calculations.

In Section 2 of this article, a short introduction for the noncommutative standard model (NCSM) is given.
Section 3 presents the calculations of the noncommutative effects on the single top quark
cross section and limit on $\theta$ from current measured single top production rate.
Finally, Section 4 concludes the paper.

\section{The Noncommutative Standard Model (NCSM)}

The NCSM action is obtained by replacing the ordinary
products in the action of the classical Standard Model by the Moyal products
and then matter and gauge fields are replaced by the appropriate
Seiberg-Witten expansions. The action of NCSM can be written as:
\begin{eqnarray}
S_{NCSM} = S_{fermions} + S_{gauge} + S_{Higgs} + S_{Yukawa},
\end{eqnarray}
This action has the same structure group $SU(3)_{C}\times
SU(2)_{L}\times U(1)_{Y}$ and the same fields number of coupling
parameters as the ordinary SM. The approach which has been used in
\cite{Madore1},\cite{Madore2},\cite{Madore3},\cite{Madore4} to
build the NCSM is the only known approach that allows to build
models of electroweak sector directly based on the structure group
$SU(2)_{L}\times U(1)_{Y}$ in a noncommutative background. The
NCSM is an effective, anomaly free, noncommutative field theory
\cite{Martin2},\cite{Martin3}. We just consider the fermions. The
fermionic part of the action in a very compact way is:
\begin{eqnarray}
S_{fermions} = \int d^{4}x
\sum_{i=1}^{3}\left(\bar{\widehat{\Psi}}^{(i)}_L\star
(i\widehat{\DS} ~\widehat{\Psi}^{(i)}_L)\right) + \int d^{4}x
\sum_{i=1}^{3}\left(\bar{\widehat{\Psi}}^{(i)}_R\star
(i\widehat{\DS} ~\widehat{\Psi}^{(i)}_R)\right),
\end{eqnarray}
where $i$ is generation index and $\Psi^{i}_{L,R}$ are:
\begin{eqnarray}
\Psi^{(i)}_L = \left(
                 \begin{array}{c}
                   L^{i}_{L} \\
                   Q^{i}_{L} \\
                 \end{array}
               \right)
~,~\Psi^{(i)}_R = \left(
                 \begin{array}{c}
                   e^{i}_{R} \\
                   u^{i}_{R} \\
                   d^{i}_{R}
                 \end{array}
               \right)
\end{eqnarray}
where $L^{i}_{L}$ and $Q^{i}_{L}$ are the well-known lepton and
quark doublets, respectively. The Seiberg-Witten maps for the
noncommutative fermion and vector fields yield:
\begin{eqnarray}
\widehat{\psi} &=& \widehat{\psi}[V] = \psi -
\frac{1}{2}\theta^{\mu\nu}V_{\mu}\partial_{\nu}\psi +
\frac{i}{8}\theta^{\mu\nu}[V_{\mu},V_{\nu}]\psi + O(\theta^{2}),\nonumber \\
\widehat{V_{\alpha}} &=& \widehat{V_{\alpha}}[V] = V_{\alpha} +
\frac{1}{4}\theta^{\mu\nu} \{\partial_{\mu}V_{\alpha} +
F_{\mu\alpha}, V_{\nu}\} + O(\theta^{2}),
\end{eqnarray}
where $\psi$ and $V_{\mu}$ are ordinary fermion and gauge fields,
respectively. Noncommutative fields are denoted by a hat. For a
full description and review of the NCSM, see
\cite{Madore1},\cite{Madore2},\cite{Madore3},\cite{Madore4}.

\section{The Noncommutative Corrections to the $t$-Channel Cross Section}

The $q_{1}(p)\rightarrow W(q)+q_{2}(k)$ vertex in the NCSM up to the
order of $\theta^{2}$ can be written as
\cite{mojtaba}:
\begin{eqnarray}\label{vertex}
\Gamma_{\mu,NC} &=& \frac{g V_{tb}}{\sqrt{2}}[\gamma_{\mu} +
\frac{1}{2}(\theta_{\mu\nu}\gamma_{\alpha}+\theta_{\alpha\mu}\gamma_{\nu}+
\theta_{\nu\alpha}\gamma_{\mu})q^{\nu}p^{\alpha} \\
\nonumber
&-&\frac{i}{8}(\theta_{\mu\nu}\gamma_{\alpha}+\theta_{\alpha\mu}\gamma_{\nu}+
\theta_{\nu\alpha}\gamma_{\mu})(q\theta
p)q^{\alpha}p^{\nu}]P_{L}.
\end{eqnarray}
where $P_{L} = \frac{1-\gamma_{5}}{2}$ and $q\theta p\equiv
q^{\mu}\theta_{\mu\nu}p^{\nu}$. This vertex is similar to the
vertex of $W$ decays into a lepton and anti-neutrino \cite{Iltan}.
However, one should note that due to the ambiguities in the SW maps
there are additional terms in the above vertex. Since they will not affect the results, we have ignored them\cite{Ruckl}.

After some algebra the noncommutative corrections to the squared matrix element
for the $t$-channel process ($u(p_{1})+b(p_{3})\rightarrow
d(p_{2})+t(p_{4})$) is as follows:
\begin{eqnarray}\label{matrixelement}
\overline{|\mathcal{M}_{NC}|^{2}} &=&
\frac{4G_{F}^{2}m_{W}^{4}|V_{tb}|^{2}|V_{ud}|^{2}}{(q^{2}-m_{W}^{2})^{2}} \times
\{ q_{i}p_{3j} [(s^{2}+u^{2})-m_{t}^{2}(s+u)] \\ \nonumber
     &-& m_{t}^{2}[sp_{2i}p_{4j}+tp_{3i}p_{4j}-up_{1i}p_{4j}
+m_{b}m_{t}q_{i}p_{4j}] \}\theta_{ij}
\end{eqnarray}
where, $m_{t}$ is the top quark mass, $m_{b}$ is the $b$-quark mass
and $m_{W}$ is the mass of $W$-boson.$V_{tb},V_{ud}$ are the CKM
matrix elements.
In Eq. \ref{matrixelement}, the contributions coming from $\mathcal{O}(\theta^{2})$
and higher have been ignored as well as the masses of light quarks.
$s,t,u$ are the Mandelstam variables.

\begin{figure}
\centering
   \includegraphics[width=10cm,height=7cm]{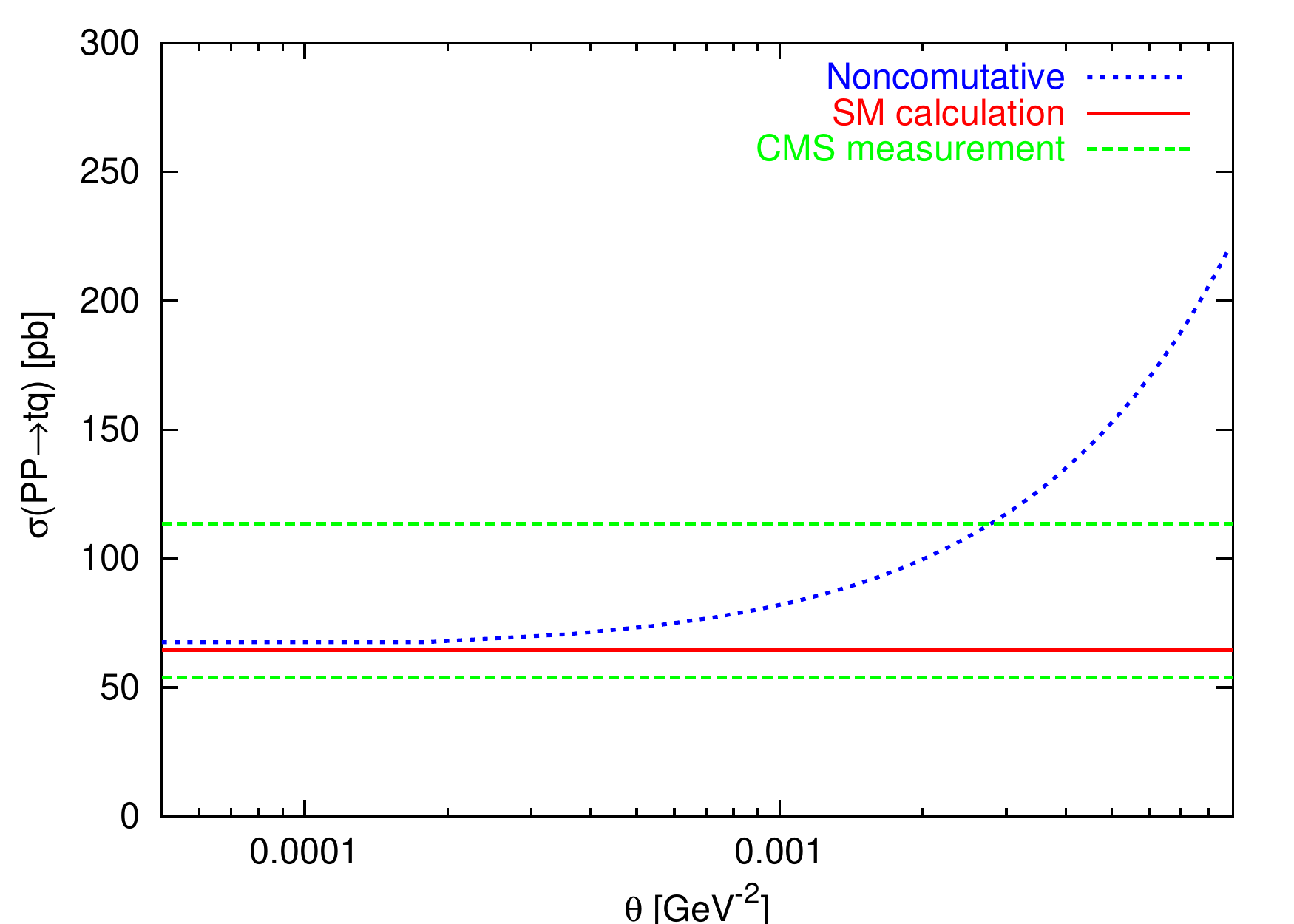} \\
  \caption{The $t$-channel single top quark production cross section
  as a function of noncommutative parameter $|\vec{\theta}|$ as well
  as the experimental precision band from the recent LHC
  measurement. The solid red line is the standard model
  value for the signel top cross section.}\label{cs}
\end{figure}

The total cross section of the $t$-channel signel top production in
proton-proton collisions at the LHC is given by:
\begin{eqnarray}
\sigma = \sum_{a,b}\int dx_{1} \int dx_{2}
f_{a}(x_{1},Q^{2})f_{b}(x_{2},Q^{2}) \hat{\sigma}_{ab}
\end{eqnarray}
where $\hat{\sigma}_{ab}$ is the partonic level cross section for the
process $u+b\rightarrow t+d$. The calculation is performed at
$Q=m_{W}$. $f_{a}(x,Q^{2})$ are the parton distribution functions. CTEQ6 \cite{cteq} is used as for the proton parton distribution
functions. Fig.\ref{cs} shows the dependence of the
single top cross section as a function of noncommutative parameter
$\theta$ as well as the experimental precision band. The solid line is the
standard model value of the single top cross section. According to Fig.\ref{cs},
the noncommutativity has constructive effect on the cross section.

Using an integrated luminosity of 36 pb$^{-1}$ collected with the CMS detector at the LHC, the value of the single top cross section found to be $83.6 \pm 29.8
(stat.+syst.) \pm 3.3(lumi.)$ pb \cite{singletop}. This measurement is consistent with the standard
model expectation.

Comparing the CMS experimental results on single top
cross section measurement with the single top cross section in the
noncommutative space-time, we get an upper value on the noncommutative
parameter $\theta$ of $\mathcal{O}(0.001)$ GeV$^{-2}$. If we assume $|\vec{B}|=1$, this limit can be
translated into $\Lambda$ and leads to $\Lambda \gtrsim 100$ GeV.
However, for smaller values of $|\vec{\theta}|$, the noncommutative corrections get smaller and smaller.
Since the LHC experiments are able to measure the $t-$channel cross section with high precision
with larger amount of data \cite{cmstdr}, this limit can be higher. For example, a measurement with $10\%$ uncertainty
leads to the bound of $\Lambda \gtrsim 500$ GeV.

\section{The Noncommutative Effect on the Spin Correlation}

One of the important features of $t$-channel single top
quark production is the large
polarization for a suitable choice
of spin quantization axis \cite{OptimalBasis}.
Because of the large mass of the top quark, it decays before
hadronizations via weak interactions and no
hadronic bound state can be formed.
On the other hand, in the SM because of the $V-A$ (Vector-Axial) structure of the
top quark couplings to the $W$ boson,
the decay products of a polarized top quark have a particular
structure of angular correlations.
The angular distributions of the decay produtcs of the top quark
have the following form \cite{alphas}:
\begin{eqnarray}
\frac{1}{\Gamma} \frac{d\Gamma}{d(\cos\theta_i)}=
\frac{1}{2}
\Bigl( 1+\alpha_i\cos\theta_i \Bigr).
\label{dGamma}
\end{eqnarray}

where $\theta_i$ is defined as the angle between the momentum of the $i$th decay product
and the top quark spin quantization axis in the top quark rest frame.
The $\alpha_i$  coefficients are called \textit{ spin correlation coefficients}
and shows the degree of correlation
with the spin direction of the top quark.
In the leptonic decay of the top quark ($t\rightarrow l^{+}+\nu_{l}+b$), with $\alpha_{l} = 1$, the charged lepton
is maximally correlated with the top quark spin direction. Please notice that $\alpha_{\nu} = -0.32$ and $\alpha_{b} = -0.40$.

In \cite{mojtaba}, the effects of noncommutative space-time on the charged
lepton spin correlation coefficient $\alpha_{l}$ has been calculated. We showed that depending on the value of
the noncommutative characteristic scale $\Lambda$, $\alpha_{l}$ can deviate significantly
from its SM value.

\begin{figure}
\centering
   \includegraphics[width=10cm,height=7cm]{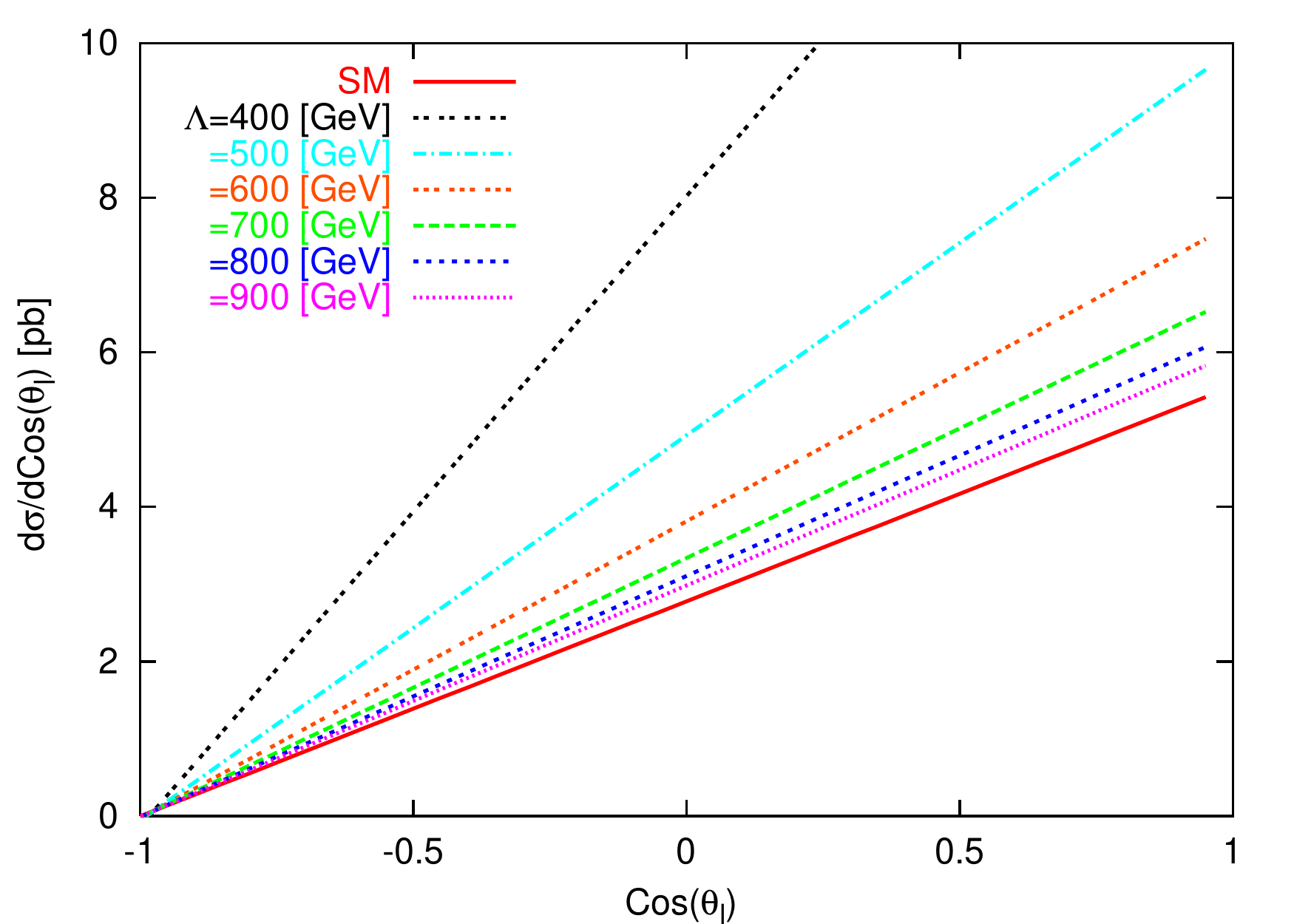} \\
  \caption{The angular correlation in $t$-channel single top quark production for the SM case with different values of the noncommutative scale $\Lambda$ at the LHC.}\label{theta}
\end{figure}

In the $t-$channel single top quark production ($u+b\rightarrow d+t$),
the direction of the spectator jet ($d-$type quark) is optimal
for the spin quantization axis \cite{OptimalBasis}. More than $96\%$
of the top quarks are produced with spin directions aligned with the momenta of the
$d-$type quark in the top quark rest frame.
In the leptonic decay of the top quark, the charged lepton
has the strongest correlation with the top quark spin, $\alpha_{l}=1$.
Accordingly, the largest
correlations appear for the case of measurement of the angle
between the charged lepton and the spectator jet in the top qurak rest frame.
Fig.\ref{theta} shows the distribution of the
cosine of the angle between the charged lepton momentum and
the $d-$type quark momentum in the top quark rest frame for the
SM case and in the noncommutative SM with different values of the noncommutative
scale $\Lambda$. According to Fig.\ref{theta}, the angular distribution is
significantly sensitive to the noncommutaitve space-time when $\Lambda \lesssim 1$ TeV.

\begin{figure}
\centering
   \includegraphics[width=10cm,height=7cm]{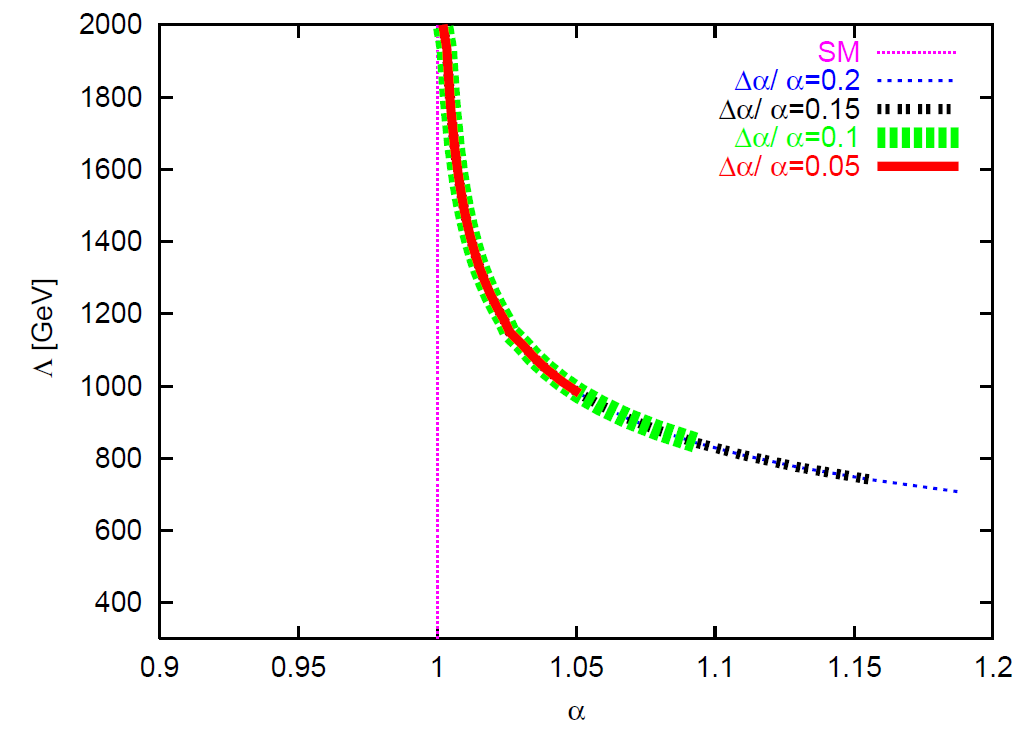} \\
  \caption{The lower limits on $\Lambda$ in GeV assuming various
relative uncertainties on the measurement of $\alpha_{l}$.}\label{error}
\end{figure}

A bin by bin precise comparison of the angular distribution with real data could provide
much better limit on $\Lambda$ with respect to the limit obtained from the total cross section
measurement. Fig.\ref{error} depict the lower bound on the noncommutative characteristic scale
$\Lambda$ in GeV as a function of the charged lepton spin correlation coefficient. In this Figure,
the lower limit on $\Lambda$ has been shown depending on different uncertainties which the 
LHC experiments could measure. For example, in Fig.\ref{error}, the end point of the thick red 
curve is corresponding to the lower limit on $\Lambda$ axis assuming the relative uncertainty on $5\%$
on $\alpha_{l}$. Table \ref{erroralpha} shows the lower limit on $\Lambda$ in GeV assuming various
relative uncertainties on measurement of $\alpha_{l}$. As you can see, the best lower limit that we can achieve 
via spin correlation from single top events is $\Lambda \gtrsim 980$ GeV.

\begin{table}
\begin{center}
\begin{tabular}{c|c}\hline
  Relative Uncertainty $\Delta\alpha/\alpha$ &  Lower limit on $\Lambda$ in GeV \\ \hline  
  $5\%$ & 980 \\
  $10\%$ & 844 \\
  $15\%$ & 742 \\
  $20\%$ & 708 \\ \hline
\end{tabular}\caption{\label{erroralpha} The lower limit on $\Lambda$ in GeV assuming various
relative uncertainties on measurement of $\alpha_{l}$.}
\end{center}
\end{table}

\section{Conclusions}

In conclusion, the noncommutative effects on the single top
quark production cross section at the LHC are very small for most
of the parameter space (less than $\mathcal{O}(0.001)$ GeV$^{-2}$).
Therefore, it seems
that there is not much hope of determining possible noncommutative
effects directly from the single top quark cross section.
However, the distribution  of the cosine of the angle between
the charged lepton momentum from the top decay and the momentum of the
spectator quark in the top quark rest frame is
highly sensitive to the noncommutaitve space-time when $\Lambda \sim 1$ TeV.
A precise comparison of the angular distribution with real data could provide
much better limit on $\Lambda$. We find the lower limit on $\Lambda$ depending
on different relative uncertainties on measurement of $\alpha$, if the
experiments could measure $\alpha$ with the precision of $5\%$, the lower 
limit on $\Lambda$ will be 980 GeV.


\end{document}